\documentclass{article}    

\usepackage{graphicx}

\title{\textbf{Observed Correlations in \mbox{Minkowski Space}}}  
\author{Richard Mould\footnote{Department of Physics and Astronomy, State University of New York, Stony Brook,
\mbox{New York} 11794-3800; http://nuclear.physics.sunysb.edu/ \~{}mould}}  
\date{}    
   
\begin{document}             

\maketitle              

\begin{abstract}

      	The author has proposed five rules that permit conscious observers to be included in quantum mechanics. 
In the present paper, these rules are applied to the observation of a non-local pair of correlated particles. 
\mbox{Rule (4)} again prevents an anomalous result.  Two different kinds of relativistic state reduction are
considered, where these differ in the way that they impose boundary conditions in Minkowski space.  In response
to a problem that arises in this context, we require the Lorentz invariance of stochastic hits.  And finally,
it is claimed that the rules proposed by the author are themselves relativistically covariant with some
qualification.      

\end{abstract}

\section*{Introduction}

 		 In several previous papers \cite{RM1,RM2,RM3, RM4}, the author proposed five rules that describe how conscious
observers become engaged with quantum mechanical systems.  These rules connect solutions to Schr\"{o}dinger's
equation with observation more completely than is possible with the Born rule.  Our rule (1) replaces the Born
rule.  These rules are not simply definitions or interpretations of the symbols in the Schr\"{o}dinger solutions, for
they add synthetic content.  They place constraints on brains that play an observational role in quantum mechanics. 
The rules are given in the appendix, and a summary and discussion of them is found in another paper \cite{RM5}.  

 		These five rules imply the existence of a wider theoretical understanding of conscious brains that is yet to be
discovered.  I make no attempt to formulate a wider theory or in any way `explain' these rules.  The rules are
correct in the sense that they set quantum theoretical solutions into correspondence with well understood
laboratory experiences.  They do a more thorough job in this respect than the Born rule by itself, and to this
extent they are to be thought of as `empirical' rules whose theoretical basis is as yet unknown.  Further
discussion is found in ref.\ 5.  

	The rules have been successfully applied to the measurement of a single particle in all of the ways that one or two
observers might interact with the measuring device.  It is shown in the present paper that the rules also apply to
the measurement of two nonlocally correlated particles, where the location of each is recorded by a different
detector.  We show how a dual measurement of this kind would look in Minkowski space, where two different strategies
for imposing reduction boundaries are considered.  In the first, an Aharonov-Albert reduction is assumed
\cite{YA1}; and in the second, a Hellwig Krause reduction is assumed \cite{HK}.  We discuss the relativistic
covariance of the rules

\section*{Correlated Particles, Non-Relativistic}

	In the following, two independent detectors are represented in their ground state (i.e., prior to particle capture)
by $D_0D_0$, where first term refers to the first detector and the second term refers to the second detector.  A
sub-zero means that the detector is in its ground (no capture) state, and a sub-one means that it is in its excited
(i.e., capture) state.    The detectors do not interact with each other.  Prior to the time that a particle
interacts with either one of the detectors, the equation of state is given by
 
\begin{displaymath}
\Phi(t \ge t_i) = \psi(x_1, x_2, t)D_0D_0 
\end{displaymath}
where $\psi(x_1, x_2, t)$ is a two particle state given as a function of non-separable spatial coordinates $x_1,
x_2$ and time $t$, where $t_i$ is the initial time. 

When each particle interacts with its own detector, the Hamiltonian of the system gives rise to three new
components.  
\begin{eqnarray}
\Phi(t \ge t_0) &=& \psi(x_1, x_2, t)D_0D_0 + \psi(a, x_2, t)D_1D_0\\
&+& \psi(x_1, b, t)D_0D_1 + D_1(t)D_1(t)\nonumber
\end{eqnarray}
where the added components are equal to zero when the interaction begins at time $t_0$.  The first detector captures
the first particle at $x_1 = a$, and the second detector captures the second particle at $x_2 = b$.  Each particle is
located inside of its own detector state $D_1$, so the second order component $D_1(t)D_1(t)$ contains both
captured particles.  

The fourth component reflects that possibility that each detector has captured a particle, where the temporal
sequence of the captures is not a consideration.  We assume that the Hamiltonian in eq.\ 1 does not allow a direct
transition from the first to the fourth component, so the fourth component can only acquire amplitude following a
second order transition that goes through the second or third component. 

	Now suppose that both detectors have been observed from a time before the beginning of the interaction.  The
initial state function of the system is then given by
\begin{displaymath}
\Phi(t \ge t_i) = \psi(x_1, x_2, t)D_0\underline{B}_0D_0\underline{B}_0 
\end{displaymath}
where the first $\underline{B}_0$ is the conscious brain state of the first observer looking at the first ground
state detector, and the second $\underline{B}_0$ is the conscious brain state of the second observer looking at the
second ground state detector.  The underline of a brain state means that it is conscious. 

	The moment the particles begin to interact with the detectors, the Hamiltonian of the system will give rise to two
other terms.  
\begin{eqnarray}
\Phi(t \ge t_0) &=& \psi(x_1, x_2, t)D_0\underline{B}_0D_0\underline{B}_0 + \psi(a, x_2,
t)D_1B_1D_0B_0\\ &+& \psi(x_1, b, t)D_0B_0D_1B_1\nonumber
\end{eqnarray}
where the second and third components are zero when the interaction begins at time $t_0$, and increase in time.  As
before, the second component represents the possibility that the first detector captures the first particle while
the second detector remains in its ground state, and the third component represents the possibility that the second
detector captures the second particle while the first detector remains in its ground state.  The brain states in the
second and third components are \emph{not} conscious states as indicated by the absence of an underline.  These are
called \emph{ready brain states}, which means that they are neither conscious nor unconscious.  They occupy a
stand-by status that has no classical counterpart.  A ready state is one that will become conscious the moment the
component is stochastically chosen in accordance with rules (1 \& 3) in the appendix.   

Notice that eq.\ 2 does not contain a fourth component like eq.\ 1.  \mbox{Rule (2)} requires that any newly created
component (in eq.\ 2) contains ready brain states, and  rule (4) does not allow a transition to go from a
ready brain state to another ready brain state of the same observer.  Since there is no direct Hamiltonian
connection from the first component to a possible fourth component in eq.\ 2, it follows that a fourth component
cannot be reached by \emph{any} route at this stage of the interaction.  Therefore a fourth component does not appear
in that equation.  As previously stated, rule (4) is a selection rule that affects the behavior of the physical
system by forbidding transitions of a certain kind - in this case, a second order transition to the $4^{th}$
component.  

	This is another case in which rule (4) spares us an anomalous result.  The fourth component of eq.\ 1 represents
the possibility that both detectors have captured a particle, where the temporal sequence of these captures is not a
consideration.  However, if there is an observer present at both detectors, then the question of the temporal order
of capture (in the given Lorentz frame) can be answered unambiguously.  It will be a matter of record that either
the first detector was the first to capture a particle or the second detector was the first to capture a
particle\footnote{Simultaneous capture in some Lorentz frame is always a possibility, but a problem is avoided by
the way the Minkowski boundaries of the reduction are defined in the section on the Hellwig-Kraus strategy.}. 
Therefore, a fourth component cannot be stochastically chosen prior to the stochastic choice of either component two
or three.  So rule (4) correctly modifies the brain's Hamiltonian in the presence of conscious observers.  

	Suppose that the second component in eq.\ 2 is the first to be stochastically chosen, corresponding to a capture of
the first particle at $x_1 = a$ at time $t^a_{sc}$.   Then from rule (3) we will get 		
\begin{equation}
\Phi(t \ge t^a_{sc}) = \psi(a, x_2, t)D_1\underline{B}_1D_0\underline{B}_0 + D_1(t)B_1D_1(t)B_1
\end{equation}
where the second component is equal to zero at $t^a_{sc}$ and increases in time.  This might be followed by a
second stochastic hit, resulting in the second particle capture at $x_2 = b$ at time $t^b_{sc}$ .  In that case, eq.\
3 becomes					
\begin{equation}
\Phi(t \ge t^b_{sc} > t^a_{sc}) = D_1\underline{B}_1D_1\underline{B}_1
\end{equation}
by virtue of rule (3).  So a dual observation can be realized with rule (4) in effect, but only after a rule (3)
reduction resulting from a single observation.    

	It is possible that there will not be a capture of either particle.  In that case, components two or three in eq.\
2 will acquire the status of a phantom components as defined in ref.\ 1.  Current will cease to flow into them, so
their subsequent existence will be of no importance to the dynamics of the  system.  They will remain dormant
until another interaction leads to a state reduction that causes them to go discontinuously to zero, or until they go
to zero by the discharge of current to a new component.   

	It is also possible that only the second particle avoids capture.  In that case, the second component in eq.\ 3
will become a phantom, lying dormant until some other interaction causes it to go to zero.

\section*{Reduction Rules (1a) or (3)}

	In ref.\ 5 we discussed the possibility that there might be state reductions that can come about \emph{without} an
observer being present.  These are called \emph{objective measurements}, and are dealt with more thoroughly in a
separate paper \cite{RM8}.  \mbox{Rule (3)} in the appendix provides for state reductions, but this happens only
when there is an observer present.  To give an objective measurement a concrete form, we propose another rule (1a)
in refs.\ 5 and 8 (also in the appendix) that provides for objective measurements independent of observation. 
Either reduction rule (1a) or (3) requires an instantaneous global reduction of the state function.  In the
following, equations are written to accommodate either kind of reduction; so the initial state in eq.\ 1 will have
square brackets around the fourth component.  This will mean that that component is equal to zero if an observer is
present, but not otherwise.  Also for \emph{observation measurements}, brain states are understood to be attached to
each component as in eqs.\ 2, 3, and 4.  
\begin{displaymath}
\Phi(t \ge t_0) = \psi(x_1, x_2, t)D_0D_0 + \psi(a, x_2, t)D_1D_0 + \psi(x_1, b, t)D_0D_1 + [D_1(t)D_1(t)]
\end{displaymath}

As before, $t^a_{sc}$  will be the time of a stochastic hit and subsequent reduction on
the first detector at point $x_1 = a$; and  $t^b_{sc}$ will be the time of a stochastic hit and subsequent
reduction on the second detector at point $x_2 = b$.  Dropping the reference to $t_0$, the initial state will now
be written $\Phi(t^a_{sc}, t^b_{sc} > t)$, which refers to any time before either of the stochastic times
$t^a_{sc}$  and  $t^b_{sc}$.  The function $\psi(x_1, x_2, t)$ is assumed to include a description of its own initial
preparation.   There are then four separate solutions that are generated by the rules in the case of two correlated
non-relativistic particles.  
\begin{eqnarray}
\Phi(t^a_{sc}, t^b_{sc} > t) &=& \psi(x_1, x_2,t)D_1D_0 + \psi(a, x_2,t)D_1D_0\nonumber\\ 
&& + \psi(x_1, b,t)D_0D_1 + [D_1(t)D_1(t)]\nonumber\\
\Phi(t^b_{sc} > t \ge t^a_{sc}) &=& \psi(a, x_2,t)D_1D_0 + D_1(t)D_1(t)\\
\Phi(t^a_{sc} > t \ge t^b_{sc}) &=&  \psi(x_1, b,t)D_0D_1 + D_1(t)D_1(t)\nonumber\\
\Phi( t \ge t^a_{sc},t^b_{sc}) &=&  D_1D_1\nonumber
\end{eqnarray}

Figure 1 shows the temporal regions in which these four equations apply.  The diagram on the left assumes that event
\textbf{A} occurs first at time $t^a_{sc}$, and the diagram on the right assumes that event \textbf{B} occurs first
at time $t^b_{sc}$ .

\begin{figure}[t]
\centering
\includegraphics[scale=0.8]{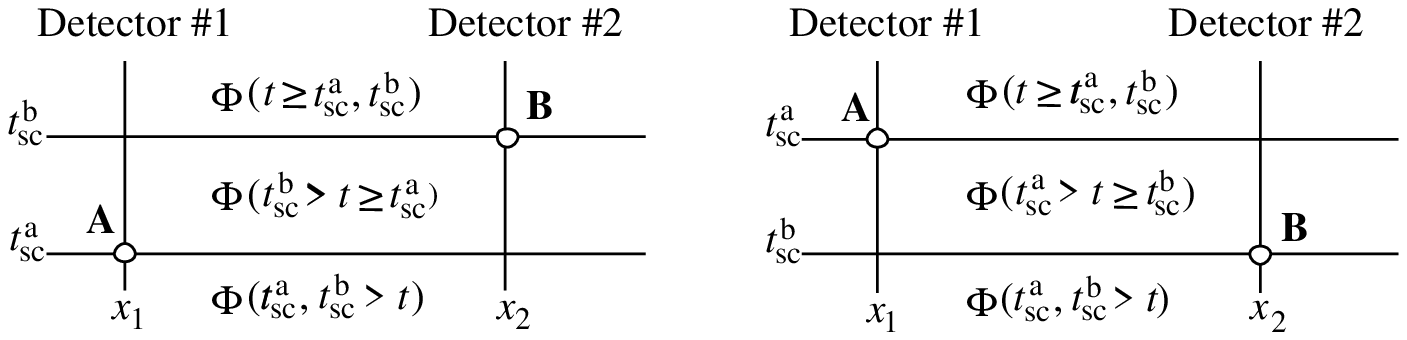}
\center{Figure 1}
\end{figure}

Remember the significance of these equations.   The solutions to Schr\"{o}dinger's equation are not themselves
grounded in observation.  They need to be supplemented by rules that tell us what happens (empirically) when specific
boundary conditions are added.  A state reduction is equivalent to adding a new boundary represented by one of the
horizontal lines in fig.\ 1, but Schr\"{o}dinger's equation cannot alone describe that process.  Therefore, the
adopted rules tell us how the Schr\"{o}dinger solutions change when a new boundary is imposed by an event such as
\textbf{A} or \textbf{B}.  The effect is to discontinuously change the solution as one moves from one region in
space-time to another.

\section*{Relativistic Reduction: \mbox{Aharonov-Albert Strategy}}

	One must decide how a reduction splits up the regions of Minkowski space, for it is not obvious how these
boundaries should be relativistically drawn when \textbf{A} and \textbf{B} are space-like events.  Different
strategies have been worked out.  Aharonov and Albert propose that a state reduction occurs along the $t = 0$ line
in each Lorentz coordinate system (ref.\ 6).   

Figure 2 shows a Minkowski map of the Aharonov-Albert solutions as witnessed by two different Lorentz observers: One
who sees event \textbf{A} occur first, and one who sees event \textbf{B} occur first.  In fig.\ 1, events
\textbf{A} and \textbf{B} on the left side are understood to be \emph{different} than \textbf{A} and \textbf{B} on
the right side.  However in fig.\ 2, \textbf{A} and \textbf{B} are the \emph{same} events seen by different Lorentz
observers.  Nonetheless, the solutions divide these two space-time regions in a very similar way.  

\begin{figure}[h]
\centering
\includegraphics[scale=0.8]{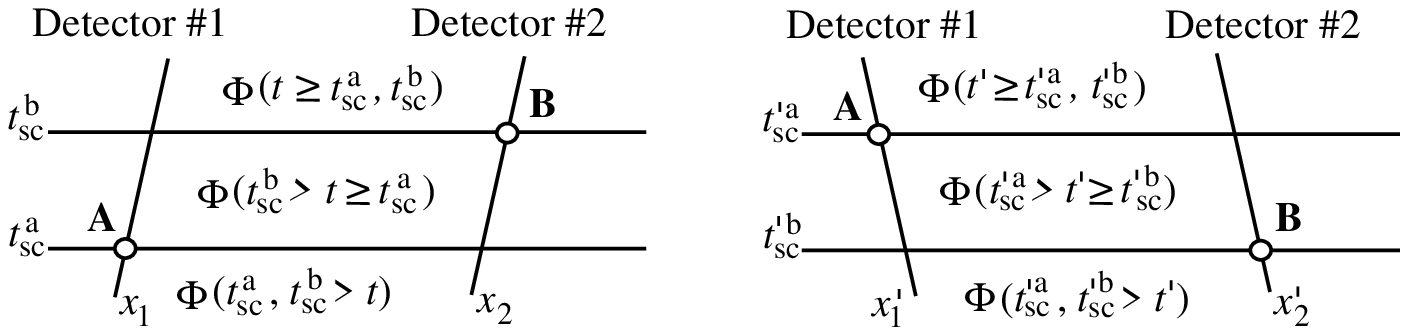}
\center{Figure 2}
\end{figure}

	The Aharonov-Albert strategy makes it impossible to imagine a general transformation that will carry a state $\Phi$
(defined at a single world point) from one Lorentz frame to another.  Consider an event \textbf{X} on the world line
of \mbox{detector \#1} just prior to   event \textbf{A} in fig.\ 2.  According to the first Lorentz observer, the
solution at event \textbf{X} will be $\Phi(t^a_{sc}, t^b_{sc}  > t)$, but according to the second Lorentz observer,
the solution at \textbf{X} will be $\Phi(t'^{a}_{sc} > t' \ge t'^{b}_{sc})$.  No local transformation can carry
$\Phi(t^a_{sc}, t^b_{sc}  > t)$ into $\Phi(t'^{a}_{sc} > t' \ge t'^{b}_{sc})$, for that crosses the boundary between
regions.  This does not deter Aharonov and Albert, for they do not try to transform a state $\Phi$ between Lorentz
frames at a single world point.  Instead, they represent reduced states as surface functionals, thereby allowing
different reductions to apply along different spacelike hypersurfaces \cite{YA2}.  This can be done in a way that
preserves the appearance of covariance.  It means that the boundary between regions is not an invariant, but is a
function of the chosen hypersurface.

\section*{The Hellwig-Kraus Strategy}

\begin{figure}[b]
\centering
\includegraphics[scale=0.8]{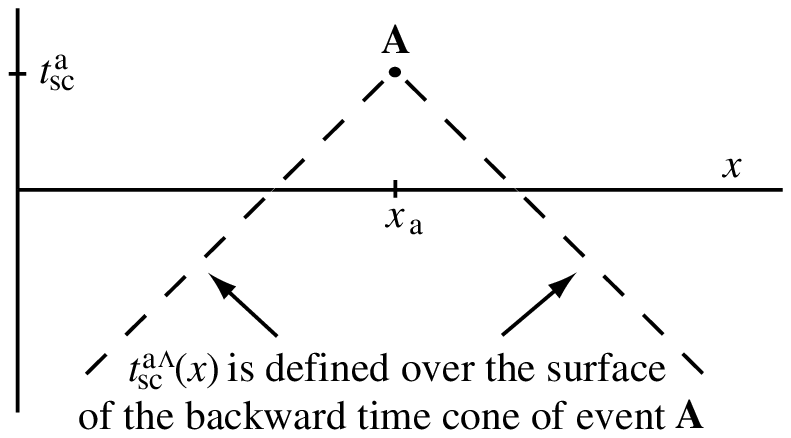}
\center{Figure 3}
\end{figure}

In ref.\ 7, Hellwig and Kraus proposed that a measurement at an event \textbf{A} gives rise to a state reduction
that takes place over the entire surface of the backward time cone of \textbf{A}.  If the reduction occurs at
the time  $t^a_{sc}$ shown in fig.\ 3, we define another time given by 
\begin{equation}
t^{a\Lambda}_{sc}(x) = t^a_{sc} - |x - x_a|/c \hspace{.5cm}\mbox{with}\hspace{.1cm} c = 1
\end{equation}
This is a function of $x$ over entire the backward time cone of \textbf{A}.  So $t > t^{a\Lambda}_{sc} $ refers to
all times $t$ after $t^{a\Lambda}_{sc}$  for any value of $x$, and this includes all events forward of the backward
time cone.  Also, $t < t^{a\Lambda}_{sc} $ refers to all times $t$ that are inside of the backward time cone of
event \textbf{A}.  At the apex we have  $t^{a\Lambda}_{sc}(x_a) = t^a_{sc} $.  Hellwig and Kraus 
require that the function that is reduced at event \textbf{A} is defined for all times $t < t^{a\Lambda}_{sc} $,
and the function that results from the reduction is defined for all times $t \ge t^{a\Lambda}_{sc}  $.

	Consider the two detector case in which there is a stochastic hit (on detector \#1) at event \textbf{A}, but there
is no hit on detector \#2.  Equation 5 is then just two solutions.
\begin{eqnarray}
\Phi(t^{a}_{sc} > t) &=& \psi(x_1, x_2, t)D_0D_0 + \psi(a, x_2,
t)D_1D_0\nonumber\\ && +\psi(x_1, b, t)D_0D_1 + [D_1(t)D_1(t)]\nonumber\\
\Phi(t \ge t^{a}_{sc}) &=& \psi(a, x_2,  
t)D_1D_0 + D_1(t)D_1(t)\nonumber
\end{eqnarray}

To impose Hellwig-Kraus boundaries on this solution, one need only change $t^a_{sc}$  to $t^{a\Lambda}_{sc}$  in the
argument of
$\Phi$.  This gives
\begin{eqnarray}
\Phi(t^{a\Lambda}_{sc} > t) &=& \psi(x_1, x_2, t)D_0D_0 + \psi(a, x_2,
t)D_1D_0\nonumber\\ && +\psi(x_1, b, t)D_0D_1 + [D_1(t)D_1(t)]\\
\Phi(t \ge t^{a\Lambda}_{sc}) &=& \psi(a, x_2, t)D_1D_0 + D_1(t)D_1(t)\nonumber
\end{eqnarray}
as shown in fig.\ 4.  The component $D_1(t)D_1(t)$ is zero at time $t^{a\Lambda}_{sc}$.

\begin{figure}[b]
\centering
\includegraphics[scale=0.8]{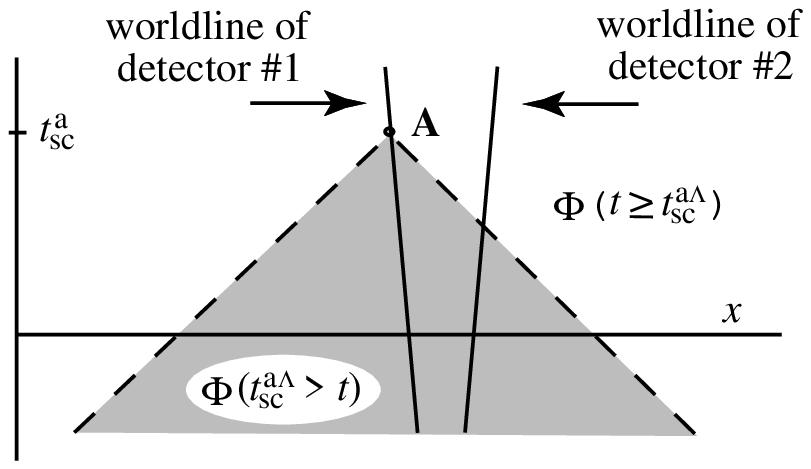}
\center{Figure 4}
\end{figure}

The first two rows in eq.\ 7 are defined in the shaded backward time cone of \mbox{event \textbf{A}}, and the third
row of eq.\ 7 is defined on the surface of the backward time cone and everywhere forward of that surface.  Both
detectors are exposed to the possibility of a stochastic hit  in the backward time cone.  But forward
of that surface, only the second detector is exposed to the possibility of a stochastic hit from the once reduced
state $\psi(a, x_2, t)$.

	Now return to the original case in which the first detector is stochastically chosen at event \textbf{A} and the
second detector is stochastically chosen at event \textbf{B}.  In this case the correct solutions are again those
in eq.\ 5, except that we impose Hellwig-Kraus boundaries.  This is done by changing  $t^{a}_{sc}$ to 
$t^{a\Lambda}_{sc}$  and  $t^{b}_{sc}$ to $t^{b\Lambda}_{sc}$  in the argument of the four solutions.   
\begin{eqnarray}
\Phi(t^{a\Lambda}_{sc}, t^{b\Lambda}_{sc} > t) &=& \psi(x_1, x_2,t)D_0D_0 + \psi(a, x_2,t)D_1D_0\nonumber\\ 
&& + \psi(x_1, b,t)D_0D_1 + [D_1(t)D_1(t)]\nonumber\\
\Phi(t^{b\Lambda}_{sc} > t \ge t^{a\Lambda}_{sc}) &=& \psi(a, x_2,t)D_1D_0 + D_1(t)D_1(t)\\
\Phi(t^{a\Lambda}_{sc} > t \ge t^{b\Lambda}_{sc}) &=&  \psi(x_1, b,t)D_0D_1 + D_1(t)D_1(t)\nonumber\\
\Phi( t \ge t^{a\Lambda}_{sc},t^{b\Lambda}_{sc}) &=&  D_1D_1\nonumber
\end{eqnarray}
In an inertial system in which event \textbf{A} precedes event \textbf{B}, the result appears in the Minkowski
diagram of fig.\ 5.  Again, for an observed measurement, brain states are understood to be attached to each component
as in eqs.\ 2, 3, and 4.

\begin{figure}[h]
\centering
\includegraphics[scale=0.8]{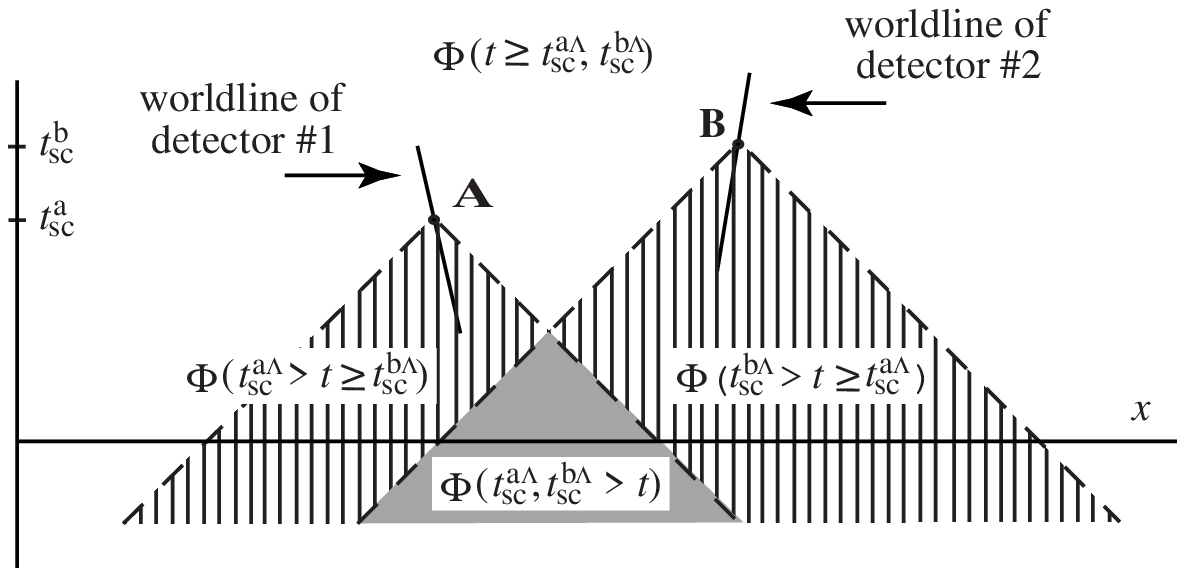}
\center{Figure 5}
\end{figure}

The immediate backward time cone of event \textbf{A} is the reduced function of event \textbf{B} (vertical lined
region on the left), and the immediate backward time cone of event \textbf{B} is the reduced function of event
\textbf{A} (vertical lined region on the right).  It is clear that boundary surfaces defined by  $t^{a\Lambda}_{sc}$
and $t^{b\Lambda}_{sc}$  are covariant, as are the four regions defined by those boundaries and occupied by the four
solutions in eq.\ 8.

The advantage of the Hellwig-Kraus strategy is that the equation of state at any world point is subject to the same
boundary conditions in any inertial system.   This means that Lorentz transformations of state are possible in
principle.  It is my inclination to favor the Hellwig-Kraus strategy for this reason.  I believe that the objections
to it raised by Aharonov and Albert can be satisfactorily answered \cite{RM10}.  This paper (ref.\ 10,  Part II)
also displays Minkowski boundaries like those in fig.\ 5, applied to a pair of 1/2 spin particles in the spin zero
\mbox{state ($J = J^2 = 0$)}.  

It is the advantage of the five proposed rules that they are just as amenable to the Hellwig-Kraus strategy as they
are to the Aharonov-Albert strategy.  One need only specify the intended boundary in the argument of $\Phi$, and the
solutions will correctly describe the intended domain in Minkowski space.  The rules of engagement proposed by the
author (appendix) give us the boundary conditions on Schr\"{o}dinger's equation, but they do not tell us which of
the above boundary types apply in Minkowski space.

\section*{Stochastic Hit Invariance}

Equation 8 does not exclude the possibility that the bracketed  $4^{th}$ component of $\Phi(t^{a\Lambda}_{sc},
t^{b\Lambda}_{sc} > t)$ is stochastically chosen before the $2^{nd}$ or $3^{rd}$ component.  That is possible under
rule (1a) absent an observerÕs ready brain state.  This means that at a certain time in the given Lorentz frame, both
particles will be simultaneously captured without any indication as to which was captured first and which was
captured second.  Suppose this stochastic occurrence happens at time $t^{ab\Lambda}_{sc}$  resulting in the equation
\begin{equation}
\Phi(t \ge t^{ab\Lambda}_{sc}) = D_1(t)D_1(t)
\end{equation}

There is a problem with eq.\ 9.  It identifies two simultaneous events in Minkowski space so there is no temporal
order placing one ahead of the other; but the events in question will have a definite temporal order in another
Lorentz frame.  This means that eq.\ 9 requires the existence of only `one' stochastic hit in the original frame, but
`two' stochastic hits in any other frame.  However, \emph{the number of stochastic hits ought to be Lorentz
invariant}; because the number of boundaries ought to be Lorentz invariant.  Equation 9 suggests that there is only
one boundary in fig.\ 5, namely the one that separates the sum of the backward time cones of (simultaneous events)
\textbf{A} and \textbf{B} from the unshaded area forward of that sum.  At the same time, there will be four
boundaries in another Lorentz frame like those shown in fig.\ 7 separating the four functions in eq.\ 8.  This same
ambiguity will exist if we used Aharonov-Albert boundaries.  

	In my mind, this difficulty provides another reason why there can be no such thing as an objective measurement. 
However, I cannot rule out the possibility that a theory that justifies rule (1a) might also solve this problem. 
That is, such a theory might provide a selection rule like rule (4) by forbidding direct transitions to second order
terms, thereby preserving the Lorenz invariance of the number of stochastic hits that are required along the
way.\footnote{If this were the case, then the bracketed component in eqs. 5, 7, and 8 would always be equal to zero.
}  I would say that any theory of objective measurement must include a rule of this kind.  So as claimed in ref.\ 8,
the choice between objective or observer measurements will depend on future theoretical developments.  It is
understood that an observer theory of measurement \emph{does} require the invariance of stochastic hits, because it
includes the selection rule (4).

\section*{Causality}

The Hellwig-Kraus scheme may suggest a causal paradox.  An observation at event \textbf{B} changes the particle
state in the backward time cone of event \textbf{A} from $\Phi(x_1, x_2, t)$ in fig.\ 4 to $\Phi(x_1, b, t)$ in
fig.\ 5, and the square modulus of the first is generally greater than the square modulus of the second.  So the
probability of event \textbf{A} is generally altered by the occurrence of an event \textbf{B} that follows
\textbf{A} (temporally) in the given inertial system.  That appears to interfere with the causal order of things,
hence a looming paradox.  However, the observer at \textbf{A} cannot possibly determine the probability of \textbf{A}
on the basis of a single observation.  An ensemble of similar experiments is necessary to find the
probability of that event. 

The threat of a causal paradox arises only when the events in question have a space-like relationship to one
another.  Only then might the observer at the \emph{single} event \textbf{A} seem to predict the observation of the
\emph{single} (future) event \textbf{B}.  But since the  observer of event \textbf{A} cannot possibly know how the
probability of \textbf{A} has been changed (or not) by a single observation, the \textbf{A} observer's
prediction of the
\textbf{B} observer's future behavior is not a possibility.   A paradox is thereby avoided.  Also see ref.\ 10, p.\
9.

\section*{Covariance of rule (1)}

	With a qualification of the word ``immediately" in rule (1a or 3) that reflects a Hellwig-Kraus boundary, the
proposed rules of engagement are valid in any Lorentz system of coordinates and are relativistically covariant. 
However, that generalization might not be clear in the case of rule (1).  

The probability per unit time is given for a single component in rule (1) by\
\begin{displaymath}
J/s = \frac{lim}{\Delta t\rightarrow 0}\frac{1}{\Delta t} \left(\frac{\Delta s}{s}\right)
\end{displaymath}
where $\Delta s$ is the change of square modulus of the component in time $\Delta t$, and $s$ is the square modulus
of the total system.   I will call the ratio $\Delta s/s$ the \emph{fractional change} of $s$ for that component.  

	Figure 6 is the Minkowski diagram of a traveling wave with the parallel lines representing lines of constant
phase.  The heavy vertical line segment is a stationary world line of length $\Delta t$ that goes between events
\textbf{a} and \textbf{b}.  The change of phase between \textbf{a} and \textbf{b} is clearly a Lorentz invariant
quantity.  Furthermore, it is reasonable to suppose that any fractional change $\Delta s/s$ of a component of the
wave function is also Lorentz invariant.  This is not known with certainty because we have no generally covariant
formulation of quantum mechanics.  However, I will state it as an assumption to be added to assumptions \textbf{A},
\textbf{B}, and \textbf{C} in \mbox{ref.\ 4}.  

\vspace{.5cm}
\noindent
\textbf{D}. Between any two events in Minkowski space, the fractional change of $s$ for any component in a
superposition is Lorentz invariant.

\begin{figure}[h]
\centering
\includegraphics[scale=0.8]{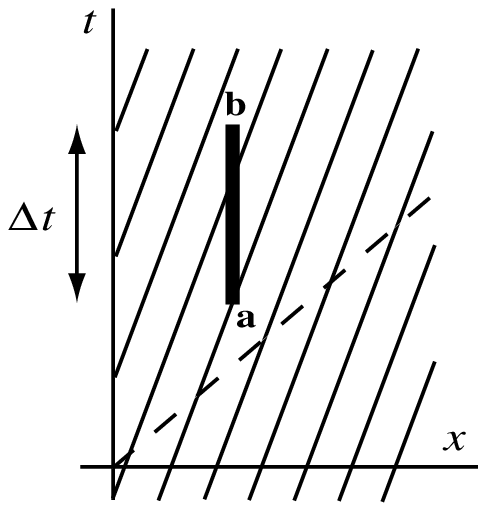}
\center{Figure 6}
\end{figure}

If event \textbf{b} is in the forward light cone of event \textbf{a} with a proper time $\Delta \tau$ between them,
then the quantity $\Delta s/s\Delta\tau$ will also be Lorentz invariant.  That is, $J/s$ for the component will be
Lorentz invariant as the system moves from event \textbf{a} to event \textbf{b}.   If, in addition, the quantity
$J/s$ is positive, then the effect of assumption \textbf{D} is to say that the probability of a stochastic hit on
the component between \textbf{a} and \textbf{b} is Lorentz invariant.  That is certainly correct.  The probability
of a stochastic hit within a given interval of proper time is the primary observable that connects a (quantum)
theory to any possible observation.  Assumption \textbf{D} is therefore empirically required of any quantum theory,
whether or not it is itself generally covariant.  

Rule (1) is stated in terms of the coordinate time $\Delta t$ rather than the proper time $\Delta \tau$.  That
presents no difficulty if it is understood that a Lorentz transformation of the probability per unit time requires a
change of the coordinate time interval $\Delta t$ (in the given inertial system) to a proper time interval   
\mbox{$\Delta\tau$  = $c\Delta t$}.  It follows that we may regard the proposed rules (in appendix below) as
relativistically covariant.

\section*{Appendix: The proposed rules}

\textbf{Rule (1)}: \emph{For any subsystem of n components in an isolated system with a square modulus equal to s,
the probability per unit time of a stochastic choice of one of those components at time t is given by
$(\Sigma_n J_n)/s$, where the net probability current $J_n$ going into the $n^{th}$ component at that time is
positive.}  

\vspace{.4cm}
\noindent
\textbf{Rule (2)}: \emph{If the Hamiltonian gives rise to new components that are not classically continuous with the
old components or with each other, then all active brain states that are included in the new components will be ready
brain states.}

\vspace{.4cm}
\noindent
\textbf{Rule (3)}: \emph{If a component that is entangled with a ready brain state B is stochastically chosen, then B
will become conscious, and all other components will be immediately reduced to zero.}

\vspace{.cm}
\begin{quote}
To reflect a Hellwig-Kraus boundary, the last sentence in rule (3) should be: ``. . . will be immediately reduced
to zero in the \mbox{space-like} region around the chosen event."
\end{quote}

\vspace{.4cm}
\noindent
\textbf{Rule (3a)}: \emph{The Hamiltonian of the brain will convert the chosen conscious state into a conscious
pulse whose width reflects the ability of the brain to resolve the conscious experience.}

\vspace{.cm}
\begin{quote}
Rule (3a), introduced in ref.\ 3, is a modification of rule 3.  It takes account of continuous nature of
brain states.  
\end{quote}

\vspace{.4cm}
\noindent
\textbf{Rule (4)}: \emph{A transition between two components is forbidden if each is an entanglement containing a
ready brain state of the same observer.}  

\vspace{.4cm}
\noindent
\textbf{Rule (5)}: \emph{If two states within a conscious pulse represent different degrees of pain, then the square
modulus of the state with the lesser pain will increase at the expense of the state with the greater pain.}

\vspace{.7cm}

\pagebreak
\noindent
\underline{Special rule in the case of an objective measurement}

\noindent
\textbf{Rule (1a)}: \emph{If the component of a superposition is environmentally decoherent, and if it is
stochastically chosen, then all of the other components in the superposition will be immediately reduced to zero.} 

\vspace{.cm}
\begin{quote}
To reflect a Hellwig-Kraus boundary, the last sentence in rule (1a) should be:  ``. . . will be immediately reduced
to zero in the space-like region around the chosen event."  
\end{quote}

\end{document}